# Eight-unit-cell electronic modulations in cuprates originating from local molecular orbitals

Zhiheng Yao[1†], Sixuan Chen[1†], Jianfa Zhao[2†], Shusen Ye[1], Weixiang Qu[1], Ning Xia[1], Yuling Dai[2], Luchuan Shi[2], Hongrui Zhang[1], Zhenqi Hao[1], Changqing Jin[2], Shuo Yang[1,3,4*], and Yayu Wang[1,3,4*]

[1]State Key Laboratory of Low Dimensional Quantum Physics, Department of Physics, Tsinghua University, Beijing 100084, China

[2]Beijing National Laboratory for Condensed Matter Physics, Institute of Physics, Chinese Academy of Sciences, Beijing 100190, China

[3]New Cornerstone Science Laboratory, Frontier Science Center for Quantum Information, Beijing 100084, China

[4]Hefei National Laboratory, Hefei 230088, China

[†] These authors contributed equally to this work.

*Corresponding author. Email: shuoyang@tsinghua.edu.cn; yayuwang@tsinghua.edu.cn

**Abstract:**

The pair density wave (PDW) state with eight-unit-cell ($8a_0$) periodicity has been widely regarded as the primary order in cuprates, yet its existence and origin remain subjects of intense debate. Using spectroscopic imaging scanning tunneling microscopy, we observe spatial modulations of the electronic states with approximately $8a_0$ periodicity in both the superconducting and insulating regimes of hole-doped $Ca_2CuO_2Cl_2$ cuprate. We find that the $8a_0$ spatial patterns are generated by the formation of molecular orbitals by doped holes, which organize into $4a_0 \times 4a_0$ plaquettes as the basic unit. Our results identify the $4a_0$ molecular orbital as the fundamental electronic building block in cuprates, while the $8a_0$ PDW represents a spatial subharmonic that emerges at sufficiently high doping.

**Main Text:**

The pair density wave (PDW) can be generally described as a superconducting state in which the Cooper pairing exhibits periodic oscillations in real space and carries a net momentum in *k*-space [1–3]. The PDW has long been conjectured theoretically to be a primary order [4–6] behind various anomalous phenomena in cuprates [7–12]. In recent years, scanning probe techniques have directly visualized the PDW state in cuprates, igniting intense interests [13–17]. In particular, scanning tunneling microscopy (STM) on optimally doped $Bi_2Sr_2CaCu_2O_{8+\delta}$ (Bi-2212) revealed a modulation with eight-unit-cell ($8a_0$) periodicity within vortex halos induced by external magnetic field [15]. Scanning Josephson tunneling microscopy, utilizing a Bi-2212 flake as superconducting tip, discovered an $8a_0$ modulation in the gap map without external magnetic field, which is considered compelling evidence for PDW [16].

Despite these advances, heated debates persist regarding the nature of the $8a_0$ PDW. First, is the PDW a long-range density wave order or is it better characterized as short-range modulations? Although the Fourier transform (FT) of gap map does reveal a $2\pi/8a_0$ wavevector, pinpointing the corresponding long-range order in real space has been difficult, with only sporadic features showing roughly $8a_0$ spacing. Second, is it the primary order driving all other phases, as suggested by many theories, or a subsidiary feature derived from another order? There are reports of PDW with the same $4a_0$ checkerboard periodicity [13,14,17], but the search for its secondary charge order by scattering probes has been fruitless. Third, how does the PDW emerge upon doping the parent Mott insulator? All prior PDW experiments on cuprates were performed near optimal doping; however, to reveal its origin, it is crucial to investigate the severely underdoped regime where pairing just starts to emerge.

To address these critical issues, in this Letter we use STM to investigate alkali-metal-doped $Ca_2CuO_2Cl_2$ (CCOC) with varied hole densities across the insulator-superconductor phase boundary. We directly detect the $8a_0$ modulations in the differential conductance ($\mathrm{d}I/\mathrm{d}V$) maps in both the superconducting and insulating samples. The primary spatial patterns, including the $8a_0$ periodicity and low-energy unidirectional order, can be explained by the formation of molecular orbital framework with $4a_0\times4a_0$ plaquette as the basic unit. We propose that the $8a_0$ modulation represents a spatial subharmonic of the $4a_0$ molecular orbital patterns, possessing a real-space origin distinctively different from a PDW order arising from finite momentum pairing.

We begin with a superconducting CCOC with hole density $p = 0.10$ and $T_c = 14$ K. A large-area topography is shown in Fig. 1(a), and the spatially averaged $\mathrm{d}I/\mathrm{d}V$ spectrum in Fig. 1(b) exhibits a superconducting gap with size around 10 meV (black arrow) and a pseudogap around 150 meV (red arrow). The superconducting gap feature is more clearly revealed by the $-\mathrm{d}^3I/\mathrm{d}V^3$ curve in the inset. The $\mathrm{d}I/\mathrm{d}V$ maps at these two characteristic energies are shown in Figs. 1(c)-(d), where stripe- and ladder-shaped molecular orbital patterns dominate the field of view (FOV), interrupted only by a small number of dark regions. FT analysis of the $\mathrm{d}I/\mathrm{d}V$ maps shown in Figs. 1(e)-(f) reveals pronounced peaks at $q \approx 1/4$ and $q \approx 1/8$ (in units of $2\pi/a_0$) along the line profiles taken through the Bragg peaks in Fig. 1(g), corresponding to real-space modulations with $4a_0$ and $8a_0$ periodicities. The $q \approx 1/4$ peak is attributed to the periodic order of plaquettes with size around $4a_0\times4a_0$, a feature well-established in previous reports on the checkerboard phase [18–27].

The observed $8a_0$ modulation matches the periodicity of the PDW in optimally doped Bi-2212 [16]. Unlike prior Josephson-tunneling study that requires a superconducting tip, here the $8a_0$ modulation is directly resolved in $\mathrm{d}I/\mathrm{d}V$ maps using a normal metal tip. To further examine its relationship with superconductivity, we analyze spatial maps of superconductivity-related quantities (Supplemental Fig. S1). Pronounced signatures of the $8a_0$ modulation are obtained from both the gap depth ($H$ = d$I$/d$V$($V_{SC}$) - d$I$/d$V$(0)) and sharpness ($D$ = -d$^3I$/d$V^3$($V_{SC}$)) maps. The consistent presence of the $8a_0$ modulation across these quantities establishes it as an intrinsic modulation of the local pairing amplitude.

To reveal the origin of the $8a_0$ modulation, Fig. 1(i) displays a series of d$I$/d$V$ spectra taken along the magenta arrow in Fig. 1(h), which crosses two dark regions in the d$I$/d$V$ map at $V_b$ = 150 mV. All d$I$/d$V$ spectra exhibit characteristic V-shaped gaps, whose depth and sharpness oscillate periodically along the linecut. The corresponding color plot in Fig. 1(j) reveals a clear $8a_0$ modulation, where the locations with small d$I$/d$V$ values corresponds to the two dark pits. This modulation thus originates from the approximately $4a_0$ lateral extent of the dark pits separating the bright plaquettes also with a typical $4a_0$ size. Real-space masking and momentum-space filtering analyses (Supplemental Fig. S2) corroborate the dominant contribution of these dark pits to the observed $8a_0$ Fourier component. Figures 1(i)-(j) also demonstrate that the $8a_0$ modulation persists over a large energy range far beyond the superconducting gap. Therefore, it is not merely a PDW, but instead a universal modulation of the overall electronic states. This $8a_0$ modulation is also observed on another FOV of the same sample (Supplemental Fig. S3).

Previous works in underdoped $Bi_2Sr_2CuO_{6+\delta}$ (Bi-2201) and CCOC have revealed that each $4a_0\times4a_0$ plaquette accommodates two doped holes on average [20,28,29]. In Supplemental Fig. S4, we show that the results reported here are consistent with this conclusion. At a nominal doping of $p$ = 1/8, a uniform distribution of doped holes would yield a dense tiling of such $4a_0$ plaquettes covering the whole FOV. At lower doping, such as $p$ = 0.10 here, incomplete plaquette coverage inevitably produces voids. Our data demonstrate that the depletion of electronic states also occurs in the unit of an entire plaquette, which creates dark pits with size around $4a_0$. This observation further strengthens our statement that the $4a_0\times4a_0$ plaquette is the fundamental building block for electronic states in cuprates [20].

We then examine an underdoped insulating CCOC with doping level $p$ = 0.05. The d$I$/d$V$ maps acquired at 30 mV and 200 mV are presented in Figs. 2(a)-(b). Compared to the $p$ = 0.10 case, this sample contains larger dark regions devoid of doped holes, reflecting the more dilute dopant distribution. The red box in Figs. 2(a)-(b) highlights a representative region with elongated plaquette chains separated by a characteristic distance around $4a_0$. A closer inspection reveals that two neighboring chains are bridged by a $4a_0$ plaquette at one end, resulting in an empty channel between them with distance around $4a_0$. These chains typically extend for 3-5 plaquettes, with the internal stripes at low energy preferentially aligned along the same direction [20]. Together with the $4a_0$ size of each plaquette, this unique configuration generates a robust, unidirectional $8a_0$ modulation perpendicular to the chain direction.

Our previous work has established that the stripe- and ladder-shaped electronic states are derived from the molecular orbitals formed by two holes on neighboring alkali dopants [20]. Although the specific structure of molecular orbital varies with energy [29–31]—consistent with the form-factor in both STM and X-ray experiments [32–35]—the spatial arrangement of the $4a_0$ plaquettes remains robust. This invariance explains the persistence of the $8a_0$ modulation across a wide energy range. To directly visualize this robustness, we present spectra and the corresponding color plot (Figs. 2(c)-(d)) taken along the red linecut in Figs. 2(a)-(b). These data reveal an $8a_0$ modulation spanning three periods over a large energy window, similar to that observed in the superconducting sample. Notably, the spectra here exhibit U-shaped gaps, in contrast to the V-shaped gap characteristic of $d$-wave superconductivity. This observation indicates that the $8a_0$ modulation is not tightly bound to well-defined $d$-wave pairing, although it will eventually lead to the modulated superconductivity.

More importantly, the internal structure of plaquettes plays a crucial role in determining how they assemble into $8a_0$ modulations. The d$I$/d$V$ map acquired at 30 mV from another FOV in the $p$ = 0.05 sample (Fig. 2(e)) reveals long, well-aligned chains separated by approximately $4a_0$ channels, resulting in a nearly unidirectional $8a_0$ modulation (Fig. 2(f)). The magenta linecut across these chains in Fig. 2(g) show that the $8a_0$ modulation coexists with the internal stripe features within each plaquette. As previously reported [20,29], the internal stripes of adjacent plaquettes tend to align along the chain direction, as a result of molecular orbital formation. Since the $8a_0$ modulation arises from the separation between plaquette chains, its direction is strongly correlated with the orientation of the internal stripes, as evidenced in Figs. 1(c), 2(a) and 2(e). Because the average spacing between internal stripes within a plaquette is around $1.3a_0$ [20,29], the $8a_0$ and $1.3a_0$ modulations should exhibit strong positive correlations along parallel directions and much weaker correlations along perpendicular directions. This is evidenced in both the superconducting and insulating samples in Supplemental Fig. S5. These spatial correlations provide a unified description of how the molecular orbital framework emerges from $4a_0$ building blocks with well-defined internal structure.

The progression from $4a_0$ plaquettes to $8a_0$ spatial modulations becomes especially apparent when comparing the $p$ = 0.05 data with that on a more dilute $p$ = 0.03 sample. Figure 3(a) shows topography of a $p$ = 0.03 sample acquired at 23 K. The sparse dopant distribution leads to relatively isolated electronic states and short chains composed of only few $4a_0$ plaquettes, as shown in the d$I$/d$V$ map at 25 mV (Fig. 3(b)) and a typical linecut along the short chain (Fig. 3(c)). At this low doping level, it is difficult to find neighboring chains with $8a_0$ separation. Figures 3(d) to (e) display the topography and d$I$/d$V$ map at another FOV of the same sample with slightly higher local dopant concentration, and Fig. 3(f) is the cut along the red line through clustered plaquettes. These data reveal the emergence of spatial separation of approximately $8a_0$ and the correlation with internal stripe patterns, which represents the genesis of $8a_0$ modulation upon doping the parent Mott insulator. With increasing hole concentration, the plaquettes progressively assemble into extended molecular-orbital network, giving rise to longer range modulations.

To support our picture for the formation of $8a_0$ modulation, we construct a minimal model based on hard-core $4a_0$ plaquettes and study it using the Markov Chain Monte Carlo (MCMC) method [36], as described in Supplementary Material Section VI. Figure 4 presents the simulation results for two representative doping levels $p$ = 0.05 and 0.10, corresponding to the two samples studied experimentally. For $p$ = 0.05, the simulated electronic potential shown in Fig. 4(a) exhibits randomly distributed bright spots, mimicking the local disorder potential induced by dopants. The close resemblance between this simulated landscape and the experimental topography supports the realism of the model. The DOS map obtained from $4a_0 \times 4a_0$ plaquettes after reaching equilibrium is shown in Fig. 4(b), where the plaquettes form short chains along either horizontal or vertical orientation. The line profiles along the corresponding FT map in Fig. 4(c) show a peak at $q$ slightly smaller than 1/8. This feature originates from the sparse distribution of plaquette chains and is consistent with experimental observations (Fig. 2). Figures 4(d) to (f) show the corresponding results for $p$ = 0.10, with well-defined Fourier peaks at both $q$ = 1/4 and $q$ = 1/8. As the $4a_0$ plaquettes occupy a larger fraction of the surface, the 1/4 FT peak becomes more pronounced. At the same time, the relative intensity of the $q$ = 1/8 peak is reduced, reflecting the decreased number of $4a_0$ voids in this higher-doped regime.

Both the experiments and simulations illustrate the assembly of $4a_0 \times 4a_0$ plaquettes into extended chains with increasing hole density. Crucially, when two adjacent chains are bridged by intermediate plaquettes, the empty channel between them is around $4a_0$ wide. Combined with the characteristic $4a_0$ width of each chain, this configuration naturally leads to an $8a_0$ spatial modulation. Upon further doping into the superconducting regime, voids with a typical size around $4a_0$ remain, which generate short-range modulations with approximately $8a_0$ separation. The presence of $8a_0$ modulation in both the superconducting and insulating samples, and its persistence across a wide energy range, indicate that it is not directly tied to superconductivity. Nevertheless, it contains the periodic modulations of superconducting properties at sufficiently high doping, so the PDW state can be viewed as a subset of the more inclusive $8a_0$ electronic modulations.

The scenario revealed here is fundamentally different from the conventional PDW picture. First, it has a pure real-space origin stemming from the self-organization of molecular orbitals of doped holes, rather than the *k*-space picture for finite momentum pairing. The $8a_0$ modulations can be clearly pinpointed in the $dI/dV$ maps, and the local unidirectional pattern is a natural consequence of the internal structure of molecular orbitals. Second, the $4a_0$ plaquette is the primary entity that emerges earlier in the phase diagram and gives birth to the $8a_0$ modulation. This contrasts sharply with proposals that the $8a_0$ PDW is the primary order that yields the subsidiary $4a_0$ charge order.

In conclusion, we demonstrate that the $8a_0$ electronic modulation in cuprates is the spatial subharmonic of the molecular orbital patterns with $4a_0 \times 4a_0$ basic units. These results indicate that the $4a_0$ molecular plaquette, each containing approximately two holes, constitutes the elementary building block for the intertwined orders in cuprates. Elucidating the nature of the molecular orbital and its impact on the phase diagram thus represents the most crucial task in solving the mechanism of superconductivity in cuprates.

*Acknowledgements*—Yayu Wang is supported by the Basic Science Center Project of NSFC (No. 52388201), the Innovation Program for Quantum Science and Technology (grant No. 2021ZD0302502), and the New Cornerstone Science Foundation through the New Cornerstone Investigator Program and the XPLORER PRIZE. Shuo Yang is supported by NSFC (grant No. 12475022) and the Quantum Science and Technology - National Science and Technology Major Project (grant No. 2021ZD0302100). The work at IOP was supported by NSFC (grant No. 12204515), the National Key Research and Development Program of China (grants No. 2022YFA1403804, 2023YFA1406001), and the Young Elite Scientists Sponsorship Program of CAST (grant No. 2022QNRC001).

*Data and materials availability*—All raw and derived data used to support the findings of this work are available from the authors on request.

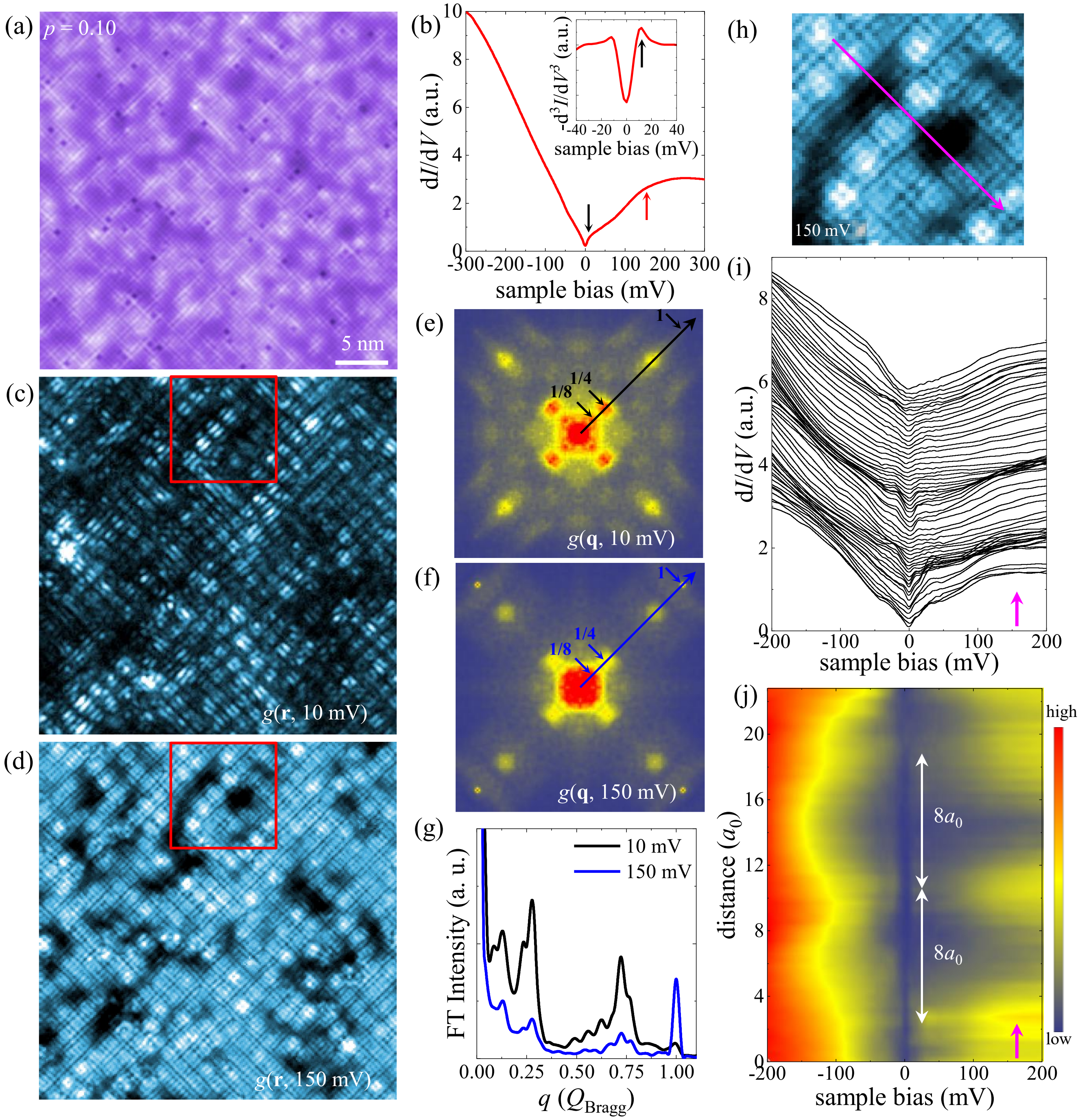

FIG. 1. Eight-unit-cell modulation in the superconducting $p$ = 0.10 sample. (a) Large-scale topographic image of the $p$ = 0.10 sample acquired at $T$ = 5 K. (b) Spatially averaged d$I$/d$V$ spectrum showing the superconducting gap and pseudogap. The inset is the -d$^3I$/d$V^3$ curve from -40 mV to 40 mV, where the superconducting gap can be clearly resolved. (c) to (d) d$I$/d$V$ maps at bias $V_b$ = 10 mV and 150 mV. (e) to (f) FT maps of (c)-(d) , revealing peaks at $q \approx 1/4$ and $q \approx 1/8$. (g) Line profiles along the Bragg directions in (e)-(f), highlighting both wavevectors. (h) Zoomed-in region of the red square in (c) and (d). (i) to (j) d$I$/d$V$ spectra and color plot along the magenta arrow in (h), showing clear $8a_0$ modulations that persist across a large energy range.

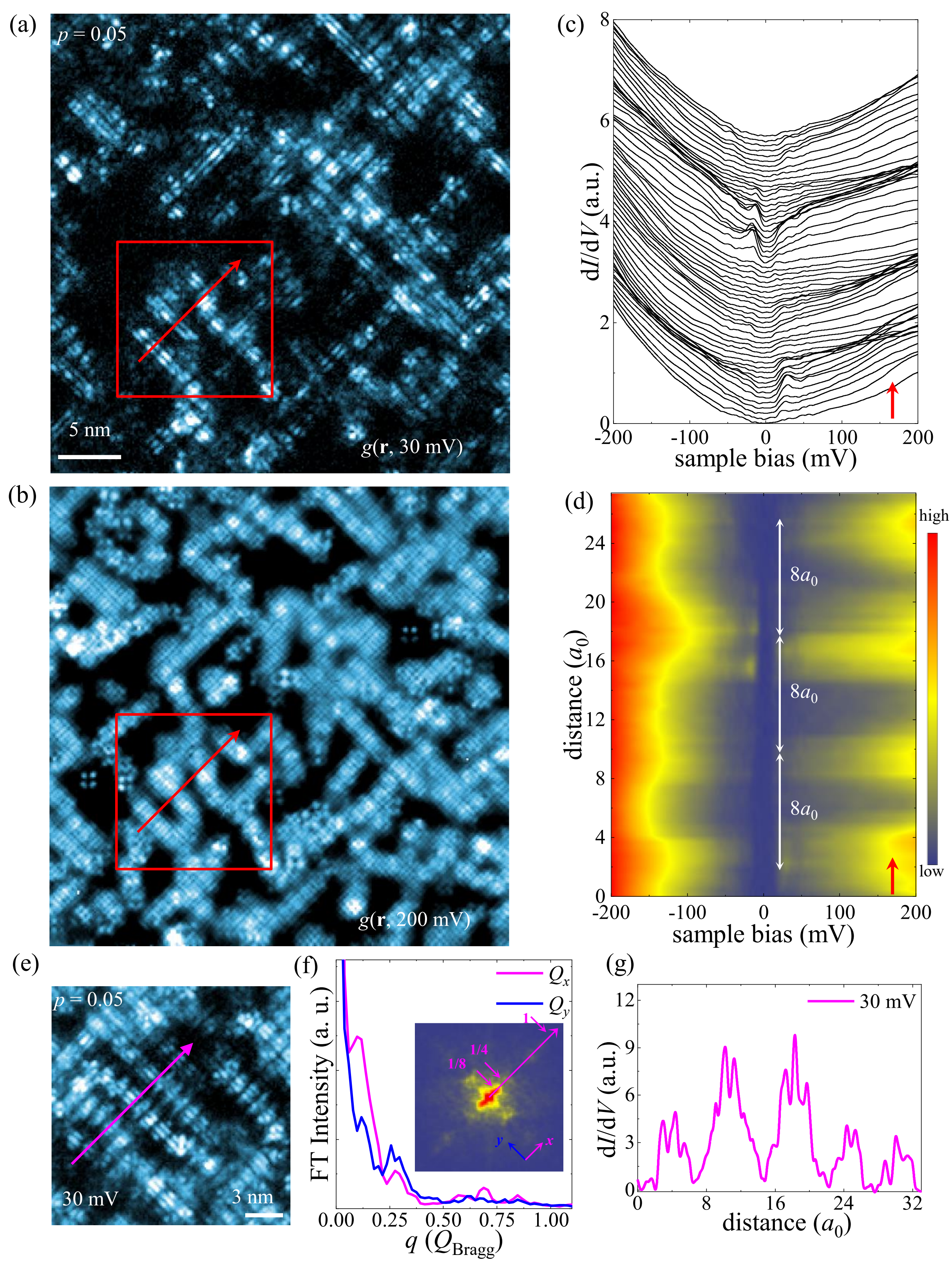

FIG. 2. Eight-unit-cell modulation in the underdoped $p$ = 0.05 sample. (a) to (b) d$I$/d$V$ maps on the $p$ = 0.05 sample at bias $V_b$ = 30 and 200 mV acquired at $T$ = 5 K, revealing the formation of molecular orbitals with energy-dependent internal structures. The region of red square highlights chain-like assemblies of $4a_0$ plaquettes separated by intervals of approximately $4a_0$. (c) to (d) d$I$/d$V$ spectra and corresponding color plot along the red arrow, exhibiting an $8a_0$ modulation over three periods. (e) d$I$/d$V$ map at bias $V_b$ = 30 mV for another FOV with unidirectional $8a_0$ modulation. (f) FT map (inset) of (e) and its line profiles along the Bragg peaks. (g) The magenta linecut revealing the coexistence of the $8a_0$ modulation and $1.3a_0$ spacing of internal stripes.

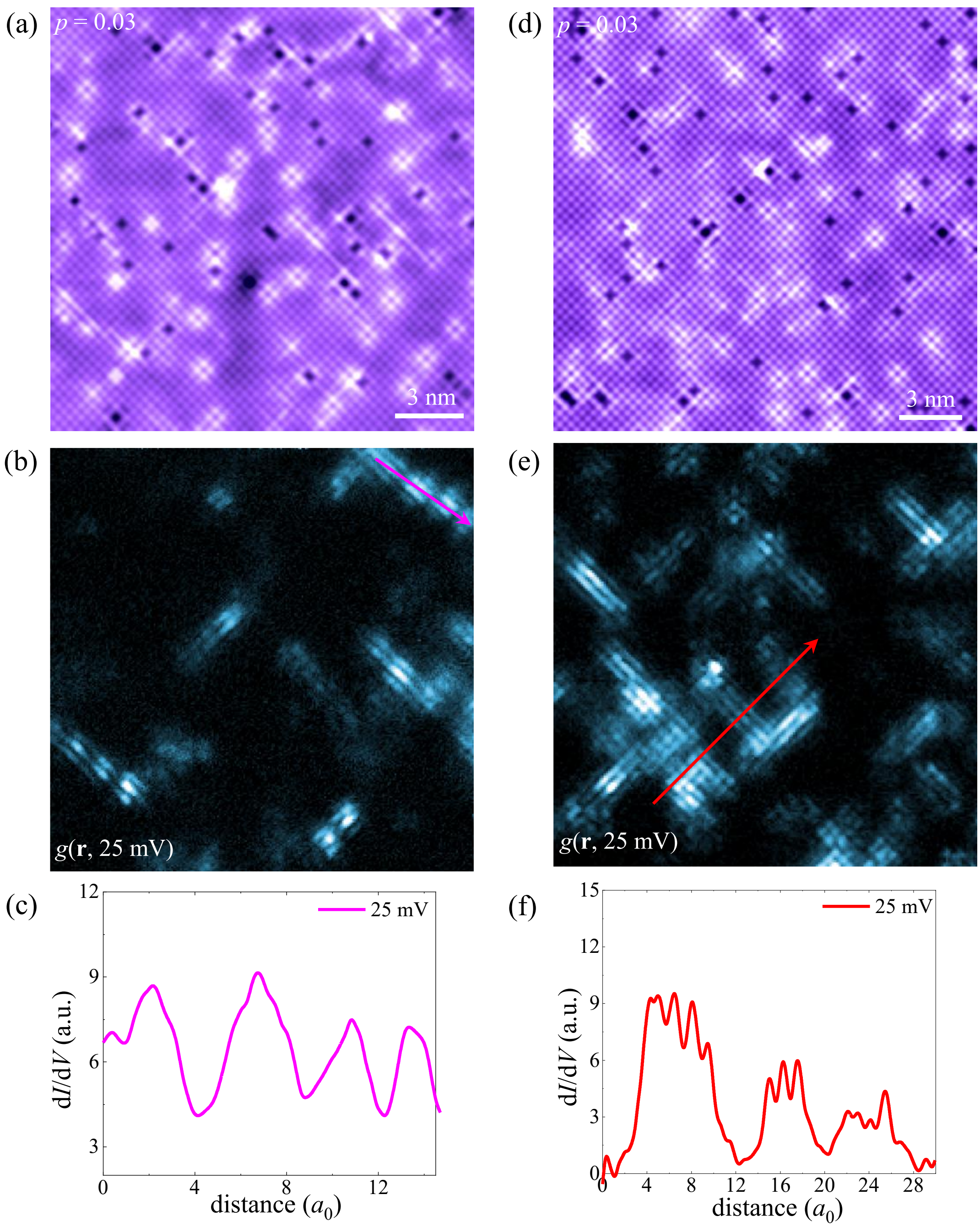

FIG. 3. Emergence of $4a_0$ plaquette and $8a_0$ modulation in the $p = 0.03$ sample. (a) Topographic image of $p = 0.03$ sample acquired at $T = 23$ K. (b) d$I$/d$V$ map at bias $V_b = 25$ mV, showing the formation of almost isolated plaquettes and short chains. (c) Linecut along the magenta arrow in (b), revealing the formation of $4a_0$ plaquette. (d) to (e) Similar dataset as (a)-(b) at another FOV of the same sample with slightly higher local dopant concentration. (f) Linecut along the red arrow in (e), revealing the emergence of $8a_0$ modulation and the internal stripes of the $4a_0$ plaquettes.

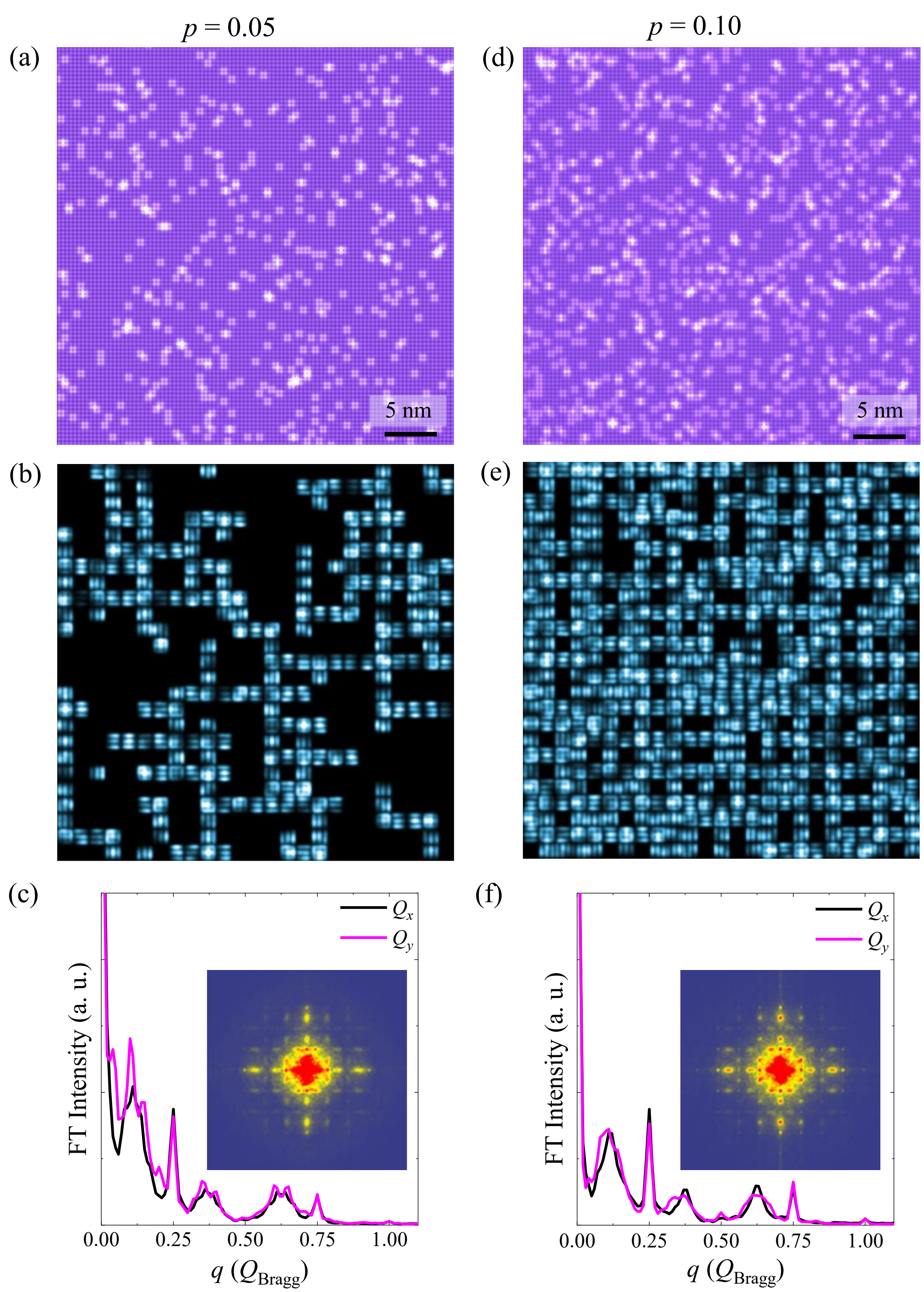

FIG. 4. Monte Carlo simulation illustrating the emergence of $8a_0$ modulations from the self-organization of $4a_0$ molecular orbitals. (a) Impurity potential generated by randomly distributed dopants for $p$ = 0.05. (b) Representative low-temperature configurations of $4a_0$ plaquettes, showing the formation of plaquette chains. (c) The line profiles along the corresponding FT (inset), revealing the dominant $4a_0$ periodicity and the accompanying $8a_0$ modulation. (d) to (f) The same set of results as in (a)-(c) for $p$ = 0.10. The plaquette chains become longer because of the increased coverage of molecular orbitals. Both the $4a_0$ and $8a_0$ periodicities are clearly observed.

Supplementary Materials for

# Eight-unit-cell electronic modulations in cuprates originated from local molecular orbitals

**Authors:** Zhiheng Yao[1†], Sixuan Chen[1†], Jianfa Zhao[2†], Shusen Ye[1], Weixiang Qu[1], Ning Xia[1], Yuling Dai[2], Luchuan Shi[2], Hongrui Zhang[1], Zhenqi Hao[1], Changqing Jin[2], Shuo Yang[1,3,4*], and Yayu Wang[1,3,4*]

**Affiliations:**

[1]State Key Laboratory of Low Dimensional Quantum Physics, Department of Physics, Tsinghua University, Beijing 100084, China

[2]Beijing National Laboratory for Condensed Matter Physics, Institute of Physics, Chinese Academy of Sciences, Beijing 100190, China

[3]New Cornerstone Science Laboratory, Frontier Science Center for Quantum Information, Beijing 100084, China

[4]Hefei National Laboratory, Hefei 230088, China

[†] These authors contributed equally to this work.

*Corresponding author. Email: shuoyang@tsinghua.edu.cn; yayuwang@tsinghua.edu.cn

**The PDF file includes:**

## Materials and Methods

Hole-doped CCOC single crystals are synthesized using a high-pressure technique. The powder of precursor is sealed in Pt cylindrical capsules and loaded into a cubic anvil high pressure apparatus at 6 GPa. Temperature is first ramped to 1300 °C, kept there for half an hour, and then cooled to 1000 °C in 6 hours. The precursors for the two lightly doped samples ($p$ = 0.03, and 0.05) are parent CCOC, $Ca(ClO_4)_2$, and NaCl with molar ratio 1:0.1:$x$, where $x$ is 0.05 and 0.10 for the $p$ = 0.03, and 0.05 samples, respectively. The hole densities are estimated by the molar ratio of Na in single crystals measured from energy dispersive X-ray (EDX) detectors. The precursors for the higher doped samples ($p$ = 0.10) are parent CCOC, $KClO_4$, and NaCl with the molar ratio 1:0.2:$x$, where $x$ is 0.08 for the $p$ = 0.10 sample, respectively. The addition of $KClO_4$ in the precursor helps enhance the hole density in this sample. Now the hole densities are estimated by summing the relative molar ratio of Na and K in single crystals measured by EDX. The susceptibility measurement finds that the $p$ = 0.10 sample is SC with $T_c$ = 14 K, which is consistent with the empirical $p$ - $T_c$ formula [1].

For STM experiments, the samples are mounted in an air bag filled with Ar atmosphere to avoid the contact of CCOC with air. The samples are cleaved in the ultrahigh vacuum preparation chamber at 77 K with base pressure lower than $1.0 \times 10^{-10}$ mbar, and immediately transferred into the STM head. The constant current mode is used to collect the topography and the differential conductance spectra are acquired by a standard lock-in method. Before the quantitative data analysis, the atomic lattice is identified by the Lawler-Fujita algorithm [2]. Fourier-transform (FT) analysis was performed after symmetrization along the Cu-O bond directions unless stated otherwise.

## Supplementary Text

### I. Pair-density modulation in superconducting $p$ = 0.10 sample

Because an $8a_0$ periodicity has been widely discussed in connection with a putative PDW state, we examine whether the observed $8a_0$ modulation in d$I$/d$V$ map corresponds to a modulation of superconductivity. A representative spectrum is shown in Fig. S1a. The depth of superconducting gap is quantified by the height difference $H = \mathrm{d}I/\mathrm{d}V(V_{\mathrm{SC}}) - \mathrm{d}I/\mathrm{d}V(0)$ as illustrated by the low-bias spectroscopy in Fig. S1b. The amplitude of the SC coherence peak is characterized by $D = -\mathrm{d}^3I/\mathrm{d}V^3(V_{\mathrm{SC}})$, which captures the curvature of the coherence peak as demonstrated in Fig. S1c. The resulting gap-depth $H$-map and the gap-sharpness $D$-map exhibit pronounced FT peaks near $q \approx 1/8$ shown in Figs. S1d-i. Together, these observations establish that the $8a_0$ modulation is accompanied by a modulation of superconducting pairing strength.

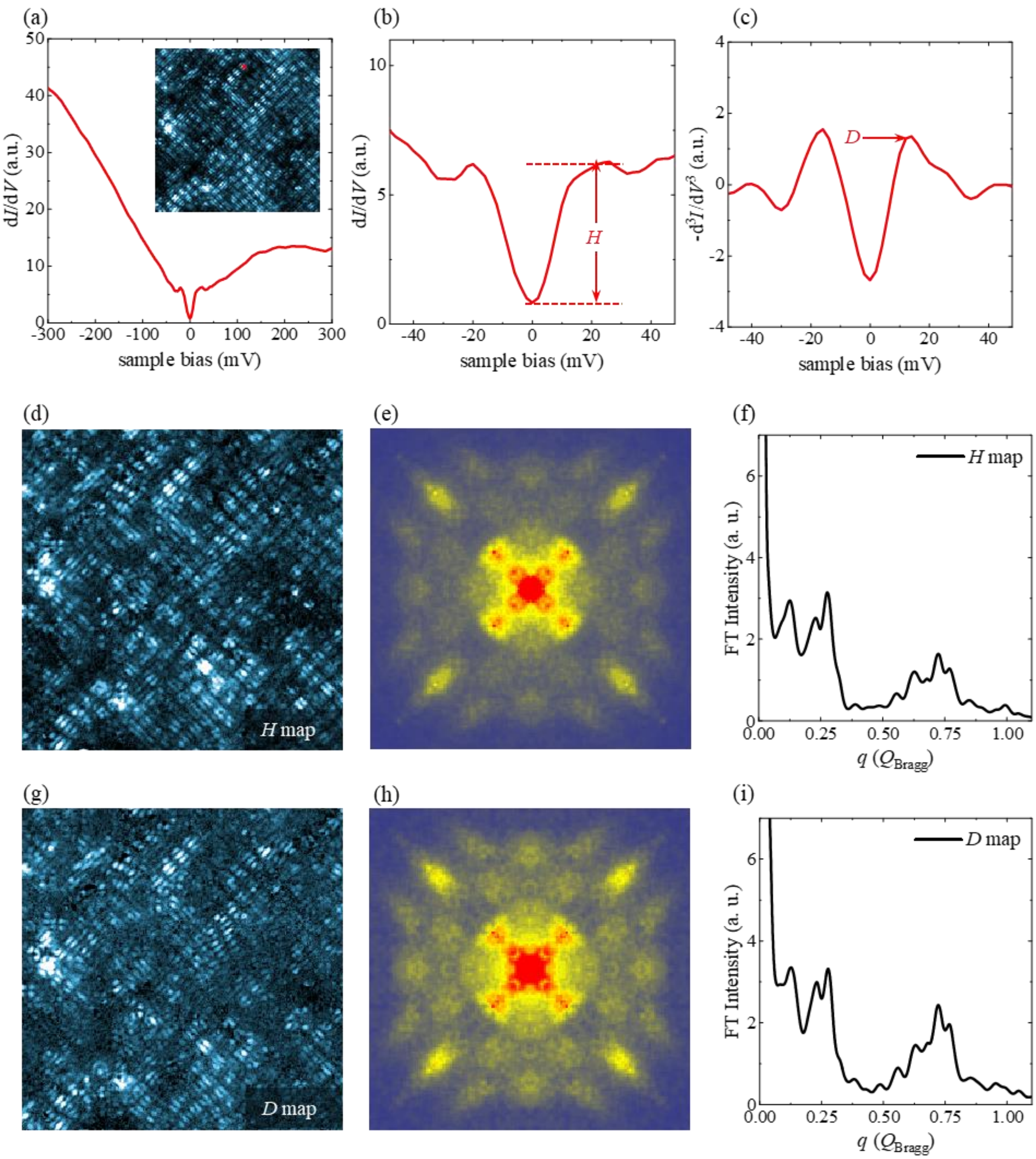

FIG. S1. Superconductivity with around $8a_0$ modulation in the $p$ = 0.10 sample. (a) A representative spectrum taken at the spots of the same FOV in Fig. 1, which is indicated by the red cross in the inset. (b) The corresponding low-bias spectroscopy of the spectrum in (a). (c) Negative of second derivative of this typical curve ($D(V) = -\mathrm{d}^3I/\mathrm{d}V^3$) with enhanced superconducting coherence peak feature. (d) The gap depth $H$-map with $H = dI/dV(V_{SC}) - dI/dV(0)$ in the superconducting $p$ = 0.10 sample. (e) to (f) The corresponding FT map and the line profile along the Bragg peak, respectively. (g) to (i) Similar dataset as (a) to (c) for the gap sharpness $D$-map with $D = -\mathrm{d}^3I/\mathrm{d}V^3(V_{\mathrm{SC}})$.

## II. Real-space contribution to $8a_0$ modulations from $4a_0$ voids in the $p = 0.10$ sample

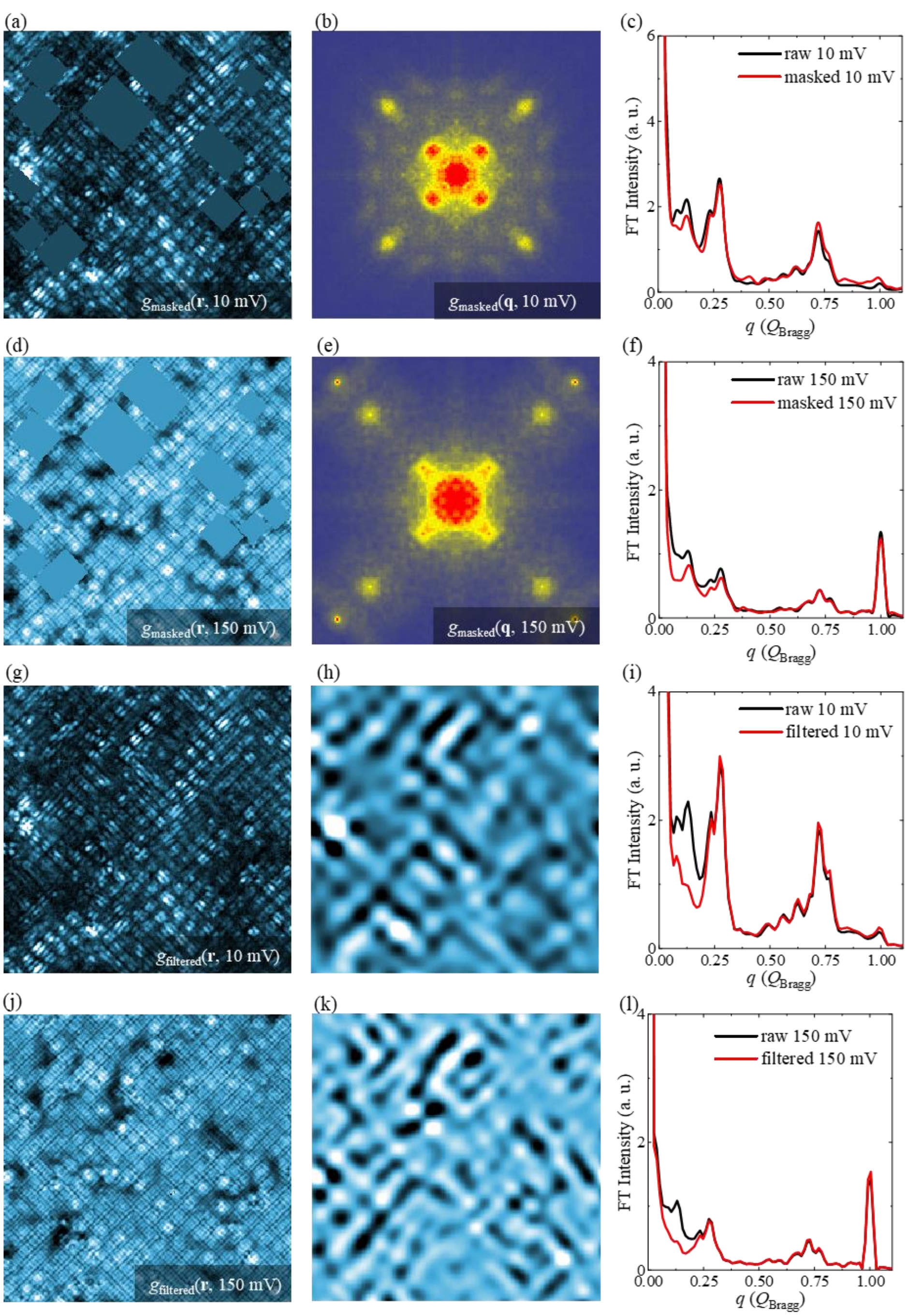

FIG. S2. Masking and filtering analyses to demonstrate the origin of $8a_0$ modulation from dark pits. (a) d$I$/d$V$ map at $V_b$ = 10 mV with the dark pit regions masked. (b) to (c) Corresponding FT map and the line profile along the Bragg peak. The black curve is the line profile in the raw data without masking. (d) to (f) Similar dataset as (a) to (c) for $V_b$ = 150 mV. (g) d$I$/d$V$ map at $V_b$ = 10 mV after filtering out the $q$ = 1/8 component in $k$-space. (h) Real-space distribution reconstructed from the band-pass-filtered $q$ = 1/8 component. (i) Line profiles along the Bragg peak in the raw and filtered map. (j) to (l) Similar dataset as (g) to (i) for $V_b$ = 150 mV.

The linecut across dark pit regions in the d$I$/d$V$ maps exhibit strong $8a_0$ oscillations that persist across a wide energy window, suggesting that these regions play a central role in generating the long-wavelength modulation. To test this hypothesis, we masked the dark pits by replacing their intensities with the spatial average at 10 mV (Fig. S2a). The FT of the masked map (Figs. S2b-c) shows a substantial reduction in the $q \approx 1/8$ peak, whereas other Fourier components remain largely unchanged. A similar suppression is observed for the 150 mV map as well (Figs. S2d-f).

To further visualize the spatial distribution of the $8a_0$ component, we band-pass filtered the FT around $q \approx 1/8$ and reconstructed the corresponding real-space signal at 10 mV (Figs. S2g-h). The resulting image shows that the $8a_0$ oscillations concentrate near the dark pit regions (Fig. S2h). This behavior persists across both low and high energies, confirming that the around $4a_0$ dark pits separating molecular orbital formation are the dominant cause for the $8a_0$ modulation.

## III. Additional field of view in the $p$ = 0.10 sample showing the $8a_0$ modulation

We have performed the same measurements on another independent field of view in the $p$ = 0.10 superconducting sample [3]. As shown in Fig. S3, it also exhibits a $q \approx 1/8$ peak in the Fourier transform of both the 10 mV and 150 mV maps, demonstrating the reproducibility of the $8a_0$ modulation across the sample.

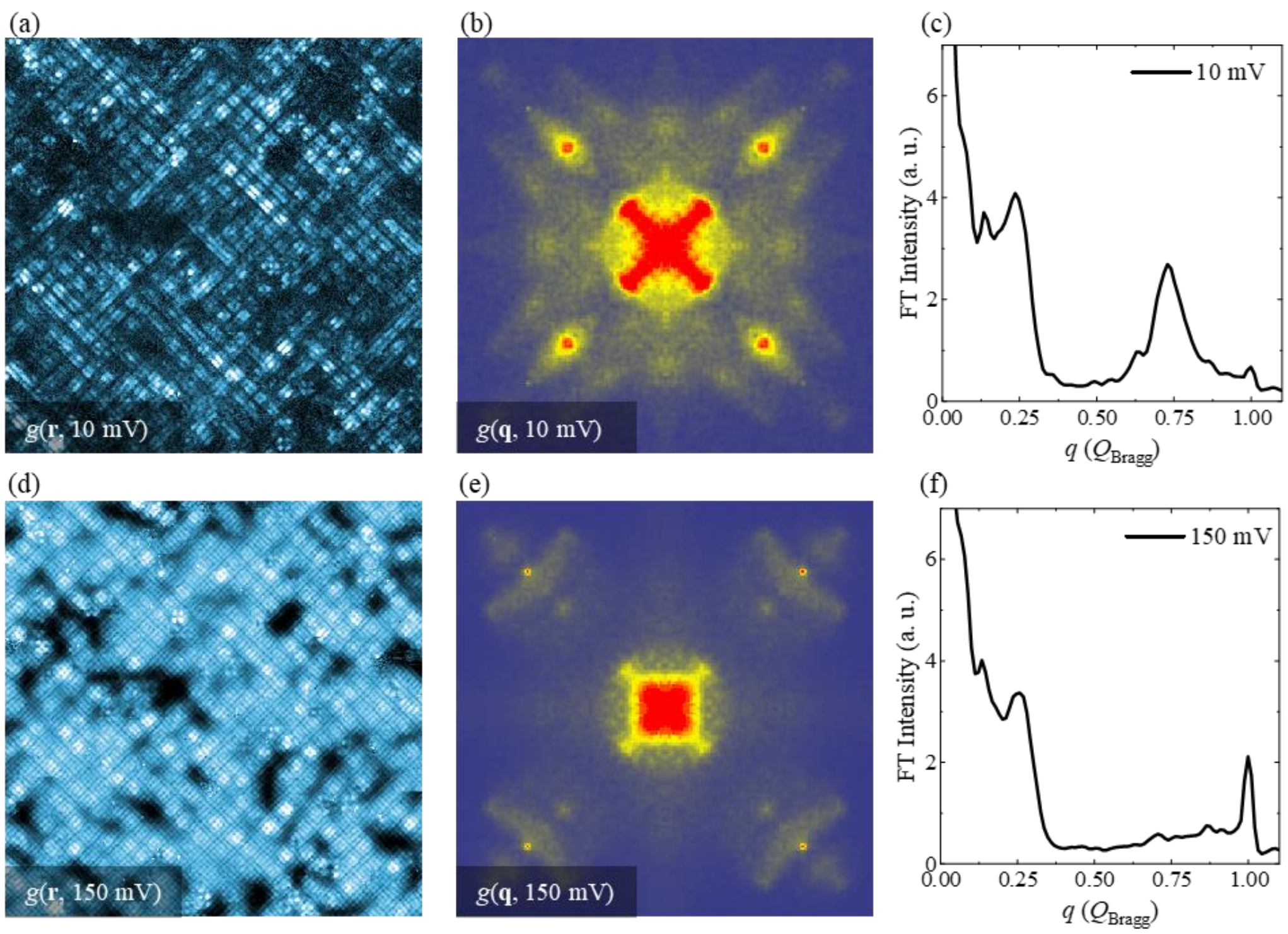

FIG. S3. Eight-unit-cell modulation in another FOV in the $p$ = 0.10 sample. (a) d$I$/d$V$ map at $V_b$ = 10 mV. (b) Corresponding FT map of (a). (c) Line profile along the Bragg peak in the FT map. (d) to (f) Similar dataset as (a) to (c) for $V_b$ = 150 mV. The dark pits with typical size around $4a_0$ are present in both d$I$/d$V$ maps, and there is a peak near $q$ = 1/8 in FT corresponding to the $8a_0$ modulation.

## IV. Estimation of the checkerboard plaquette density modulation

Previous work in Bi-2201 established that each $4a_0$ checkerboard plaquette typically hosts two doped holes [4]. We applied the same method to estimate the plaquette density in our $p = 0.10$ and $p = 0.05$ samples (Fig. S4). The detailed methodology is outlined in Ref. [4]. For the $p = 0.10$ sample, the extracted plaquette density is approximately $0.051/a_0^2$, yielding ~1.96 holes per plaquette. For the $p = 0.05$ sample, the plaquette density is approximately $0.026/a_0^2$, yielding ~1.92 holes per plaquette. These values are consistent with the expected two-hole occupancy of each $4a_0$ plaquette, and reinforce the picture that the molecular orbital formed by two holes is the fundamental real-space unit of the electronic state in cuprates.

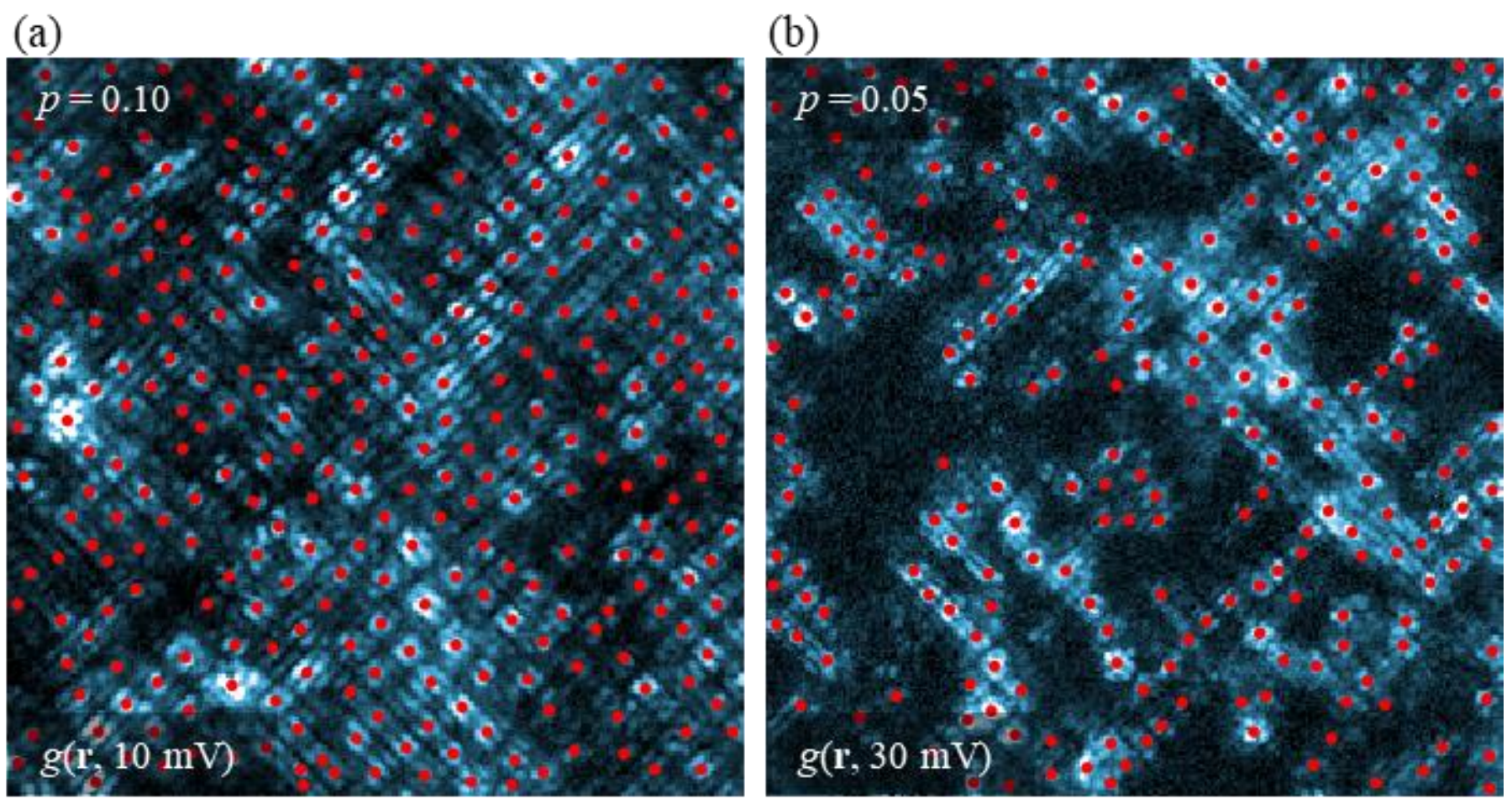

FIG. S4. Estimation of the number of holes in each plaquette. (a) d$I$/d$V$ map of the $p = 0.10$ sample at 10 mV. The red dots indicate the positions of plaquette centers. By counting the number of plaquettes, we estimate that on average there are 1.96 holes in each $4a_0$ plaquette. (b) d$I$/d$V$ map of the $p = 0.05$ sample at 30 mV. On average there are around 1.92 holes in each plaquette.

## V. Correlation between the internal stripes of plaquettes and the $8a_0$ modulations

A strong positive correlation arises between the $8a_0$ and $1.3a_0$ modulation in the local specific electronic structures, manifested in their simultaneous appearance along the same linecut (Fig. 2g and Fig. 3f). To assess whether this correlation persists across the entire FOV, we conducted a full-FOV correlation analysis, summarized in Fig. S5.

Difference of the local amplitude map [5] for $q = 3/4$ (corresponding to the $1.3a_0$ modulation) along the $x$ and $y$ directions at 10 mV in the superconducting $p = 0.10$ sample is presented in Fig. S5a. It reflects both the orientation and local strength of the $1.3a_0$ modulation. A similar analysis for the $8a_0$ modulation is shown in Fig. S5b. The correlation analysis for the amplitude map between $1.3a_0$ and $8a_0$ modulation along various relative direction is shown in Fig. S5c. It reveals a robust orientational alignment between the direction of $1.3a_0$ and $8a_0$ modulations. The correlation analysis yields consistent results for the underdoped $p = 0.05$ sample as well (Figs. S5d-f).

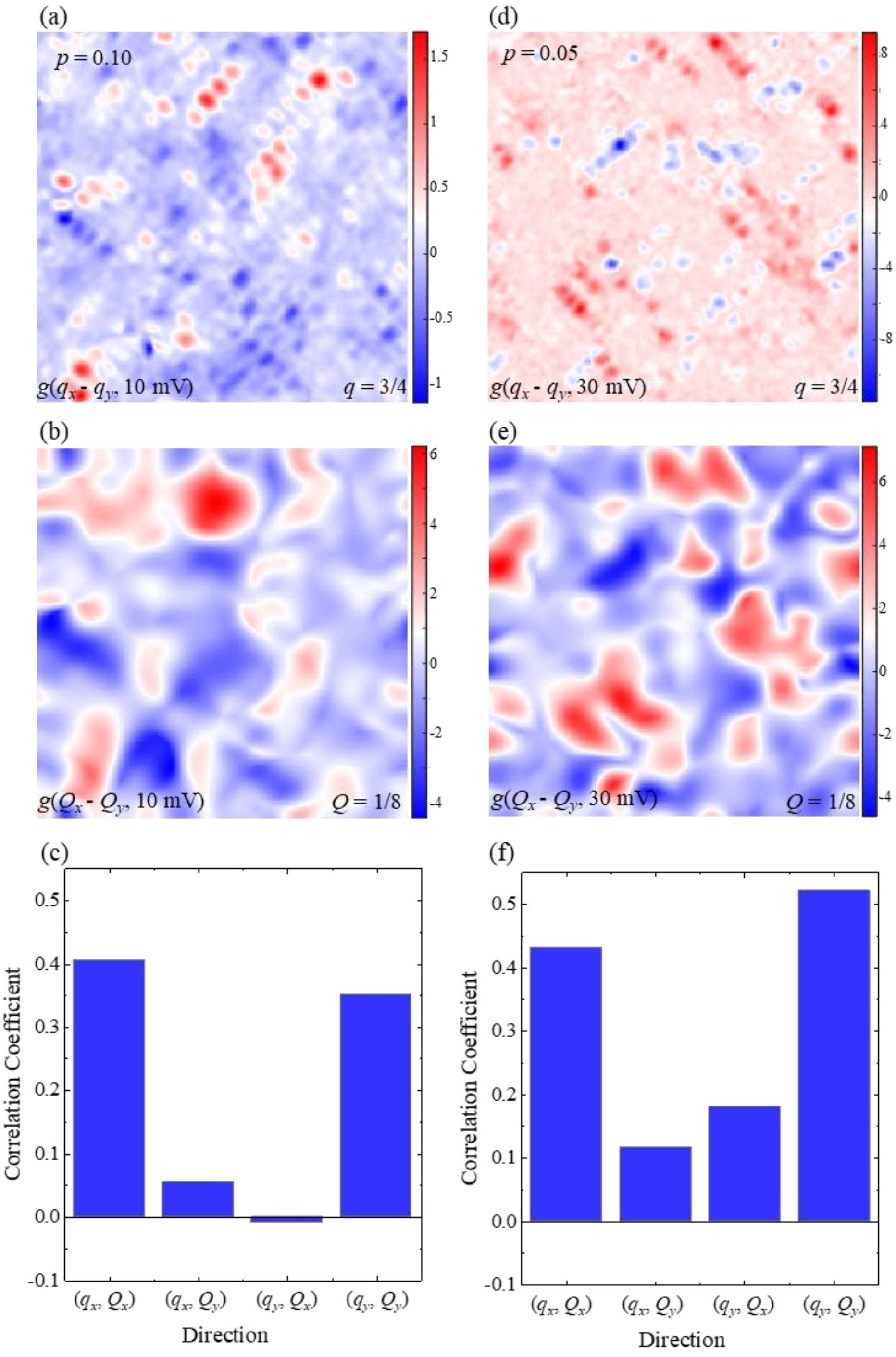

FIG. S5. Correlation between the $q = 3/4$ and $Q = 1/8$ wavevectors in the superconducting and insulating samples. (a) Difference between the amplitude map along $x$ and $y$ direction at $|q| = 3/4$ for $V_b = 10$ mV in the superconducting $p = 0.10$ sample. (b) Similar dataset as (a) for $|Q| = 1/8$. (c) Directional dependence of the correlation between the $|q| = 3/4$ and $|Q| = 1/8$ components in the $p = 0.10$ sample. (d) to (f) Similar dataset as (a) to (c) for $V_b = 30$ mV in the insulating $p = 0.05$ sample.

## VI. Markov Chain Monte Carlo simulations

The model is motivated by three robust experimental observations related to molecular orbitals. First, each $4a_0 \times 4a_0$ plaquette contains approximately two doped holes [3,4,6], allowing us to control the overall doping level by adjusting the plaquette density. Second, the internal stripe patterns of neighboring plaquettes tend to align along the same Cu-O bond direction. To describe this effect, we assign to each plaquette a discrete orientation variable $\sigma_i \in \{0, 1, 2, 3\}$ corresponding to no charge distribution, vertical stripe, horizontal stripe, and mixed stripe patterns, respectively. Specifically, when two neighboring plaquettes are aligned horizontally, the three-stripe patterns tend to orient horizontally. When they are aligned vertically, the stripes prefer a vertical orientation. At the junction between horizontal and vertical stripe regions, the plaquette favors a mixed stripe pattern. Neighboring plaquettes interact via an attractive interaction $E_{\text{int}}$, which depends on their relative stripe orientations. Third, the doped holes are subject to a dopant-induced random impurity potential $E_{\text{imp}}$.

Motivated by these observations, we construct a classical statistical model that takes the plaquette as the elementary degree of freedom. The total energy of the system is given by $E(\{\sigma_i\}) = E_{\text{imp}}(\{\sigma_i\}) + E_{\text{int}}(\{\sigma_i\})$, where $\{\sigma_i\}$ denotes a specific plaquette configuration, $E_{\text{imp}}$ describes the impurity potential acting on individual plaquettes, and $E_{\text{int}}$ accounts for interactions between neighboring plaquettes.

We first introduce the impurity potential. The Cu lattice constant is $a_0$, and Na impurities are located at the centers of four nearest-neighbor Cu sites. Each plaquette contains 16 Cu sites. A plaquette contributes to the impurity energy only when its state is nonzero. The strength of the impurity potential is assumed to be proportional to the number of Na impurities associated with the plaquette. Denoting by $m_i$ the number of Na impurities in the $i$-th plaquette, the impurity energy is given by

$$E_{\text{imp}}(\{\sigma_i\}) = -\sum_{\sigma_i \neq 0} m_i W.$$

If a Na impurity lies on the boundary or at the corner of a plaquette, it contributes $-W/2$ or $-W/4$, respectively, to each adjacent plaquette.

We next define the interaction between neighboring plaquettes. Only nearest-neighbor attractive interactions are considered, and periodic boundary conditions are imposed. For a pair of vertically adjacent plaquettes with states $\sigma_i$ and $\sigma_j$, the interaction energy is $-g_v(\sigma_i, \sigma_j)$, while for a horizontally adjacent pair it is $-g_h(\sigma_i, \sigma_j)$. The specific values of $g_v$ and $g_h$ are summarized in Fig. S6a. The total interaction energy is therefore

$$E_{\text{int}}(\{\sigma_i\}) = -\sum_{\langle i,j\rangle_v} g_v(\sigma_i, \sigma_j) - \sum_{\langle i,j\rangle_h} g_h(\sigma_i, \sigma_j),$$

where $\langle i,j\rangle_v$ and $\langle i,j\rangle_h$ denote sums over vertical and horizontal nearest-neighbor pairs, respectively. This interaction is designed to favor plaquette arrangements consistent with experimental observations.

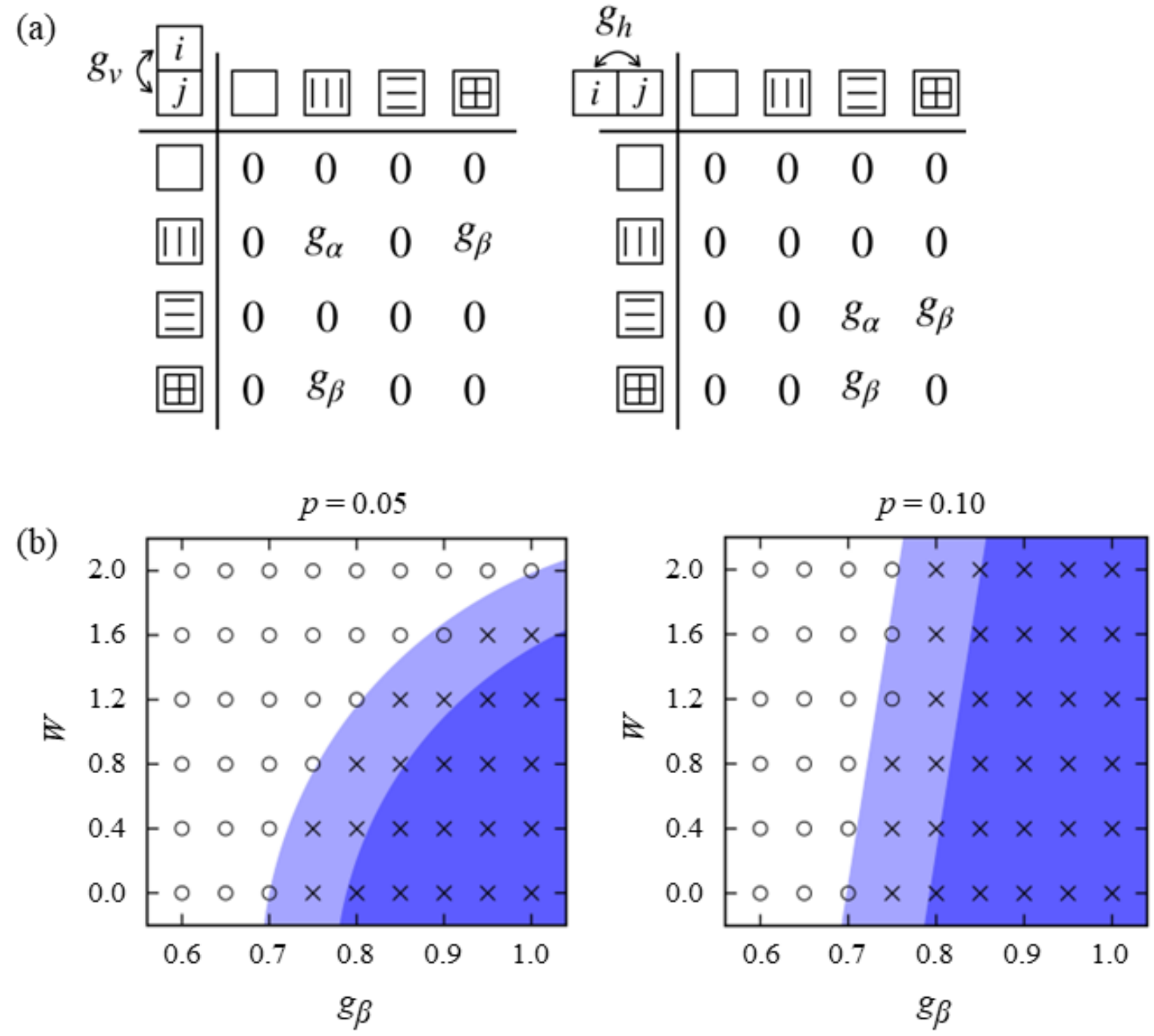

FIG. S6. Attraction interaction form and phase diagram of the MCMC simulation. (a) Strength of the attractive interaction between neighboring plaquettes. Left: vertical neighbors, right: horizontal neighbors. (b) Phase diagram showing the regions where the $q = 1/8$ peak is present (blue shaded areas) for hole doping $p = 0.05$ and $p = 0.1$, with $g_\alpha = 1$ and $T = 0.1$.

To sample the equilibrium configurations at temperature $T$, we perform Markov Chain Monte Carlo simulations using the Metropolis algorithm. The equilibrium probability distribution is given by $\exp[-E(\{\sigma_i\})/T]/Z$, where the partition function is $Z = \sum_{\{\sigma_i\}} \exp[-E(\{\sigma_i\})/T]$.

We begin by initializing the system. For a given hole doping $p$, the total number of Na impurities is fixed to $[16pL^2]$, where $[\cdot]$ denotes the integer part. These impurities are randomly placed at the centers of four Cu sites. The total number of nonzero plaquettes is set to $[8pL^2]$. Subject to the constraint on the total number of nonzero plaquettes, the initial configuration $\{\sigma_i\}$ is generated randomly. During the Monte Carlo updates, this number is kept fixed, corresponding to a fixed doping $p$.

Each Monte Carlo step consists of two types of local updates: exchanges between a nonzero plaquette and a zero plaquette, and internal state changes among the nonzero plaquettes. For each proposed update, the energy difference $\Delta E$ between the new and old configurations is calculated. Updates with $\Delta E \leq 0$ are accepted unconditionally. For $\Delta E > 0$, the update is accepted with probability $\exp(-\Delta E/T)$. The total energy is recorded every $L^2$ steps. After a sufficient number of steps, the system reaches equilibrium.

The equilibrium configurations $\{\sigma_i\}$ are then converted into real-space charge density maps. The plaquette states $\{0,1,2,3\}$ correspond to no charge distribution, vertical three-stripe pattern, horizontal three-stripe pattern, and a superposition of vertical and horizontal stripes, respectively.

To reproduce the experimentally observed spatial variation in charge intensity, we further apply a decay factor along the stripe patterns that depends on the distance to nearby impurities.

In the simulations, we fix $g_\alpha = 1.0$ and $T = 0.1$, which allows efficient sampling of configurations with distinct spatial structures. We have verified that further lowering the temperature does not qualitatively change the results. Using $g_\beta = 0.9$ and $W = 0.8$, we obtain the results for hole doping levels $p = 0.05$ and $p = 0.10$ shown in Fig. 4 of the main text, where a clear 1/8 peak is observed.

We further vary $g_\beta$ and $W$ to construct a phase diagram that indicates the presence or absence of the 1/8 peak, as shown in Fig. S6b. The 1/8 peak appears in the shaded region, which corresponds to relatively large values of $g_\beta$ and small values of $W$. In addition, compared with the case of hole doping $p = 0.05$, the 1/8 peak at $p = 0.10$ is observed over a broader range of parameters.